\def \SAIT #1 #2 {{\em Mem.\ Soc.\ Astron.\ It.\/} {\bf #1}, #2}
\def \MESS #1 #2 {{\em The Messenger\/} {\bf #1}, #2}
\def \ASTRNACH #1 #2 {{\em Astron. Nach.\/} {\bf #1}, #2}
\def \AAP #1 #2 {{\em Astron. Astrophys.\/} {\bf #1}, #2}
\def \AAL #1 #2 {{\em Astron. Astrophys. Lett.\/} {\bf #1}, L#2}
\def \AAR #1 #2 {{\em Astron. Astrophys. Rev.\/} {\bf #1}, #2}
\def \AAS #1 #2 {{\em Astron. Astrophys. Suppl. Ser.\/} {\bf #1}, #2}
\def \AJ #1 #2 {{\em Astron. J.\/} {\bf #1}, #2}
\def \ANNREV #1 #2 {{\em Ann. Rev. Astron. Astrophys.\/} {\bf #1}, #2}
\def \APJ #1 #2 {{\em Astrophys. J.\/} {\bf #1}, #2}
\def \APJL #1 #2 {{\em Astrophys. J. Lett.\/} {\bf #1}, L#2}
\def \APJS #1 #2 {{\em Astrophys. J. Suppl.\/} {\bf #1}, #2}
\def \APSS #1 #2 {{\em Astrophys. Space Sci.\/} {\bf #1}, #2}
\def \ASR #1 #2 {{\em Adv. Space Res.\/} {\bf #1}, #2}
\def \BAIC #1 #2 {{\em Bull. Astron. Inst. Czechosl.\/} {\bf #1}, #2}
\def \JSQRT #1 #2 {{\em J. Quant. Spectrosc. Radiat. Transfer\/} {\bf #1}, #2}
\def \MN #1 #2 {{\em Mon. Not. R. Astr. Soc.\/} {\bf #1}, #2}
\def \MEM #1 #2 {{\em Mem. R. Astr. Soc.\/} {\bf #1}, #2}
\def \PLR #1 #2 {{\em Phys. Lett. Rev.\/} {\bf #1}, #2}
\def \PASJ #1 #2 {{\em Publ. Astron. Soc. Japan\/} {\bf #1}, #2}
\def \PASP #1 #2 {{\em Publ. Astr. Soc. Pacific\/} {\bf #1}, #2}
\def \NAT #1 #2 {{\em Nature\/} {\bf #1}, #2}
\def \jfm #1 #2 {{\em J.\ Fluid Mech.\/} {\bf #1}, #2}

\documentstyle[twoside,epsf]{memsait}

\begin{opening}
\title{Convection in stellar envelopes: a changing paradigm}
\author{H.C. Spruit$^1$}
\institute{$^1$Max Planck Institute for Astrophysics, Garching, Germany\\}
\date{} 
\end{opening}

\begin{document}

\oddpagefooter{\sf Mem. S.A.It., Vol. ??, 1996}{}{\thepage}
\evenpagefooter{\thepage}{}{\sf Mem. S.A.It., Vol. ??, 1996}
\ 
\bigskip

\begin{abstract}
Progress in the theory of stellar convection over the past decade is reviewed. The
similarities and differences between convection in stellar envelopes and 
laboratory convection at high Rayleigh numbers are discussed. Direct numerical simulation of
the solar surface layers, with no other input than atomic physics, the equations of
hydrodynamics and radiative transfer is now capable of reproducing the observed heat flux, 
convection velocities, granulation patterns and line profiles with remarkably accuracy. These 
results show that convection in stellar envelopes is an essentially non-local process, being 
driven by cooling at the surface. This differs distinctly from the traditional view of 
stellar convection in terms of local concepts such as cascades of eddies in a mean 
superadiabatic gradient. The consequences this has for our physical picture of processes in 
the convective envelope are illustrated with the problems of sunspot heat flux blocking, the 
eruption of magnetic flux from the base of the convection zone, and the Lithium depletion 
problem. 

This text will appear in {\em Mem. Soc. Astron. It.}, 1996.
\end{abstract}

\section{Introduction}
The demands put on theories of stellar convection depend somewhat on the observables to be 
reproduced. If the aim is to reproduce the positions of stars with convective envelopes in 
the HRD, all that the theory has to provide is a single number: the mean efficiency of 
convection. This can be measured, for example, by the effective mixing-length to scale height 
ratio $\alpha$: the value that a model using the mixing length formalism would need to have 
in order to get the same position in the HRD. Equivalently, its output would be the entropy 
$S_{\rm b}$ at the base of the convection zone, as a function of the effective temperature 
$T_{\rm eff}$, surface gravity $g$ and the element abundances. The bulk of a convective 
envelope is isentropic, so that $S_{\rm b}(T_{\rm eff},g)$ is determined almost entirely by 
the properties of convection in a thin surface layer. Detailed information on this layer is 
contained in the shapes of spectral lines of a star, which depend in a 
nontrivial way on the temperature and velocity fields at the surface, as well as on the 
characteristics of the spectral line. This information has been collected for many stars 
(Nordlund and Dravins 1990, Dravins and Nordlund 1990, Gray 1982, Gray and Toner 1985), and 
provides important tests 
for theories of stellar convection. For the Sun, of course, a very much greater amount of 
detail is available. The detailed patterns of velocity and temperature are known, and 
frequency shifts of p-mode oscillations which are due to the structure of the surface layers 
are to be explained, as well as the level of excitation of the modes. 

These are all phenomena which have their cause in the uppermost layers of the convection 
zone. In addition there are manifestations of convection which have their origin in much 
deeper layers of the envelope, such as the differential rotation and the magnetic activity, 
both caused by the interaction of rotation with convective flows. 

Historically, practical descriptions of stellar convection have centered on the mixing length 
picture, the essence of which is the {\em local} approximation: one makes the assumption that 
the heat flux $\bf F$ at any point in the envelope is given uniquely by the local temperature 
gradient $\nabla T$, plus the local thermodynamic variables. Theories based on this local 
assumption can be formulated starting from different points of view. One of these is the 
original mixing length formalism (B\"ohm-Vitense 1958), but several others exist (for 
example, Gough 1977, Canuto and Mazzitelli 1991, 1992). They all lead to very similar looking 
expressions for the 
functional form of ${\bf F}(\nabla T)$, though numerical coefficients and the exponents in 
various factors in the expressions may differ. Formulations have also been devised to build 
upon these equations while making them more `nonlocal', typically by adding diffusion 
equations for some of the quantities entering the equations (Xiong 1985, Baker 1987, for 
more references see Spruit, Nordlund and Title 1990). 

Parallel to these developments, 
attempts have been made occasionally to describe stellar surface convection directly in terms 
of the dynamics of flows near the radiating surface (e.g. Namba and van Rijsbergen 1976). 
This approach remained qualitative and had little impact until numerical simulations became 
practicable in the late 70's (Nelson 1978, Nordlund, 1982). In the two decades since then, 
the initial limitations due to finite 
numerical resolution have been overcome, and the simulations have converged numerically with 
respect to the most important quantities such as the value of $S_{\rm b}$ (or alternatively 
the heat flux if $S_{\rm b}$ is given), the temperature and velocity fields at the 
surface, and the resulting spectroscoping diagnostics. The simulations are also getting 
close to reproducing the level of excitation of p-modes and the `frequency anomalies' of 
p-modes that have their origin in the outer convective layers. 

The simulations are presenting us (and have done so for more than a decade) with a picture of 
convection in stellar envelopes which differs markedly from the old `eddies in a mean 
gradient' picture. In short, they show us that stellar surface convection is {\em nonlocal} 
in an extreme way. The consequences of this change of view are remarkable, and affect many of 
our standard views of processes in convective envelopes. Before getting into these 
consequences, it is instructive to discuss briefly some laboratory experiments on convection 
at high Rayleigh numbers and related numerical simulations, since significant developments 
have take place in this field as well over the past decade.

\section{Intermezzo: Laboratory convection at high Rayleigh numbers}
Around 1985 Libchaber and collaborators (Heslot, Castaing and Libchaber 1987, Castaing et al. 
1989, Wu and Libchaber 1992, see also Tilgner et al. 1993 and references therein) performed a 
series of laboratory experiments with Helium gas in a cell at a 
temperature of 5K. By varing the pressure of the gas, the viscosity could be varied over a 
wide range, such that Rayleigh numbers up to $Ra=10^{14}$ were reached. Though this is still 
smaller than the Rayleigh numbers in a convective stellar envelope, these results have 
significant astrophysical relevance. The flow in the cell was not visualized directly, but a 
probe in the cell recorded temperature fluctuations as a function of time. At high Rayleigh 
numbers, these records showed that the temperature was `almost always' at a constant level 
midway between that of the top and bottom surfaces (call this $\Delta T=0$). From time to 
time upward or downward spikes in temperature occurred; the width of these spikes decreased 
with increasing $Ra$. If this record is represented as a 
histogram showing the relative frequency of a temperature fluctuation as a function of its 
value, one obtains a narrow distribution around $\Delta T=0$, compared with much wider 
distributions at lower $Ra$, and a non-gaussian shape. Such distributions (which also occur 
in other contexts) were called `hard turbulence', a name suggesting 
interpretation in terms of a special form of turbulent cascade. It turns out 
however that the spikes, and the narrow distribution, were due to a nonlocal effect: the 
spikes are caused by the passage across the probe of threads of buoyant fluid that were 
created at the top and bottom surfaces of the cell. Cooling (heating) of a thin boundary 
layer at the top (bottom) surface causes narrow lanes or threads of fluid to descend (rise), 
maintaining their identity as they cross the cell. The flow is driven entirely (in the limit 
of infinite $Ra$) by these threads. The fluid in between is passive, it has zero temperature 
fluctuation and is neutrally buoyant. It flows towards the boundaries to replace the mass 
flowing in the threads, rather than being driven there by convective buoyancy. Convection in 
a cell at high $Ra$ is therefore as non-local as it can possibly be: the heat flux is 
carried by flows that maintain their identity from one surface to the other. Since these 
experiments were done, numerical simulations have been done for the same conditions, and the
flow in these results can be visualized in all its details (Kerr, 1996). These simulations 
reach only Rayleigh numbers of the order $10^7$, but agree closely with the experiments, and 
confirm the `thready' nature of the flow.

Once this picture for convection in a cell is understood, it is easy to find approximate 
physical models that reproduce accurately the observed dependences on Rayleigh number 
(Castaing et al. 1989, see also Kerr 1996; for an astrophysical application see Spruit, 
1992). Such models focus on what happens in the boundary layers, a key ingredient being the 
{\it geometry} of the flow. Turbulent spectra and cascades, two- or three-point correlations 
play a lesser role in these pictures since these weakly nonlocal, algebraic, concepts do not 
do sufficient justice to the highly nonlocal and geometric nature of the problem. To what 
extent do turbulent cascades and mixing still play a role? The ascending and descending 
threads of buoyant fluid are surrounded by layers of strong shear, where one expects to find 
instabilities causing eddies and mixing, and causing the threads to broaden as they 
travel towards the other end of the cell. 
This form of turbulence and attendant mixing must be present, but the experiments 
also show that its importance for the transport of heat {\em decreases} with increasing $Ra$. 
This is shown by the fact that the amplitude of the temperature fluctuations in the bulk of 
the flow (outside the threads) decreases as a power law, $Ra^{-0.15}$, without noticeable 
breaks (see figure 13 in Castaing et al. 1989). 

In addition, turbulence is produced in the 
top and bottom boundary layers, and plays a role in the precise dependence of the Nusselt 
number (heat flux) on $Ra$. This turbulence is due to the shear in the boundary layers, 
because of the no-slip conditions in the experiments. In the astrophysical case, the 
conditions are free-slip and much less turbulence is expected due to the boundary layers (see 
section \ref{tur}). 

Putting it somewhat simplistically, at very large $Ra$ the flow acquires a simple geometric 
order, and looks less and less like a turbulently mixed flow. 
The flow is extremely nonlocal, nonisotropic (thin threads and boundary layers) and 
inhomogeneous. It is poorly represented by its spatial power spectrum. A mixture of neutrally 
buoyant fluid mixed with narrow threads may well have a spectrum that looks like a power law. 
But there will be many density fields with exactly the same power spectrum but entirely 
different geometrical structure. One does not expect that such flows will behave like the 
real thing. Somewhat besides the present topic, I note that this is not restricted to the 
present  context of convection in a cell. It is already apparent in experiments on 
homogeneous isotropic turbulence. Numerical simulations of such turbulence (Siggia 1983, Kerr 
1985, Vincent and 
Meneguzzi 1994, Porter et al.\ 1994) show that in a statistically steady state the smallest 
scales occur, at least in part, in the form of narrow vortex tubes whose length is of the 
order of the scale at which the flow is forced. Even though the spectrum of such flows is 
close to Kolmogorov, this shows that there are connections between the smallest and the 
largest scales that are not even approximately covered by a local turbulent cascade model.

Given that the boundary layers are of such overriding importance for everything that happens 
in the cell, it is natural that simplified models as well as numerical simulations should 
focus on the processes in these layers. For recent numerical studies in the context of the 
laboratory convection experiments, see Kerr (1996), Kerr et al. (1996) Cattaneo et al. 
(1991), DeLuca et al. (1990) and Rast and Toomre (1993). In astrophysics, 
numerical studies with a clear focus on the physics of the surface boundary layer have been 
done as soon as they were technically feasible. 

\section{Stellar surface convection}
In some respects, convection in a stellar envelope is similar to the laboratory case. The 
thermal boundary layer at the surface is again quite thin (20 km for the Sun), and the narrow 
threads descending from this layer are very similar. Some things are very different, however. 
The most important of these is {\em stratification}. In the laboratory cell the net density 
gradient due to gravity plays no role, and the processes taking place at the top and bottom 
surfaces are entirely symmetrical. In the solar convection zone, the gas density at the 
bottom is $10^6$ times as high as at the surface. This destroys the symmetry between top and 
bottom in a fundamental way. Because of the high density at the bottom, the heat flux can be 
carried with a very small temperature contrast. The familiar mixing length estimate, for 
example, predicts temperature differences between upward and downward moving fluid of a few 
degrees, i.e. $\Delta T/T\sim 10^{-6}$, whereas at the surface $\Delta T/T$ is of the order 
0.3. Fluid rising from the base is therefore uniform in potential temperature (the 
temperature after adiabatic expansion to a given pressure) to one part in $10^5$ when it 
arrives at the surface. The buoyancy of the upward threads generated at the base is 
negligible compared with that of the downward threads generated at the surface. As a 
consequence, for computations of convection at the surface, the entropy of the upward flowing 
fluid may be considered uniform for all practical purposes (Nordlund 1982, 1985c, Nordlund 
and Stein 
1990). This would not be the case if significant mixing between upward and downward flows 
were to take place throughout the convection zone, as assumed in the mixing length (and in 
any other local) picture. The laboratory experiments mentioned above as well as the numerical 
simulations, however, show that such mixing is very limited. 

The first consequence of the extreme asymmetry between top and bottom is that upward moving 
fluid is {\it isentropic}, hence neutrally buoyant, all the way from the base of the 
convection zone 
to the thermal boundary layer at the top. It follows that we can not say anymore that the 
flow is driven by buoyant bubbles moving up from below. All driving (in terms of actual 
forces) is due to {\em cooling at the surface}. The gently rising isentropic upflow gets 
exposed to the cold universe in a thin layer near optical depth unity. The extreme 
temperature dependence of the opacity makes the cooling happen even faster than it would have 
done otherwise: as the fluid cools, it becomes more transparent, and cools even faster. The 
cooled fluid has strong (negative) buoyancy, and collects in a downward flowing lane. On its 
way down the lanes very soon break up into threads or `raindrops'. The upward flow of the hot 
fluid serves to replace the fluid `condensing' at the top. The entropy contrast of the 
downflows, the horizontal and vertical velocities, the length scales of the upwellings are 
all determined by the physics happening in the surface boundary layer.

\subsection{What is the role of turbulence?}
\label{tur}
The picture sketched above is essentially nonlocal and highly geometric. To what extent can 
concepts from turbulence theory, with its focus on local cascades and functions in Fourier 
transform space be applied here? Formally, one can take a spatial power spectrum of the 
density field, for example, at some depth, take its temporal mean, and call this the 
turbulent spectrum. Though 
such spectra contain useful information, a traditional interpretation of terms of 
turbulent cascases would be meaningless, given that most of the structure in the spectrum 
is due to the downward moving threads that were generated at a very different depth 
(namely, the surface). The spectrum has lost the local meaning which it is implicitly or 
explicitly assumed to have in theories of turbulence. 

Though a description of the threads in terms of turbulent cascades does not make much sense, 
turbulence does play some role. In the numerical simulations of laboratory convection, it is 
found (e.g. Kerr, 1996 and references therein) that the fluid between threads is actually 
turbulent in a more traditional sense. Since this fluid is neutrally buoyant, the turbulence 
in it is probably mechanically driven by the 
shear at the surfaces and at the edges of the threads. Turbulence produced by the shear flow 
at the edges of the threads is not of direct consequence for determining the heat flux, nor 
for the patterns observed at the surface, since it is created {\em after} the flow has left 
the cooling surface. But the degree of mixing in these shear layers is important for the 
amount of surrounding fluid the threads capture by entrainment, and therefore the speed with 
which they reach the bottom. This is discussed somewhat further in section \ref{li}. At this 
point, it is important to take into account the experimental finding that the width of the 
threads, in spite of entrainment, {\em decreases} with Rayleigh number instead of increasing 
as one might have expected of the basis of the simple `higher $Ra$ means more turbulence 
means more mixing' line of argument. 

It has often been remarked that the cellular pattern of granulation looks remarkably smooth 
in spite of the very high Reynolds number of the flow. Why should small scale turbulence be 
so weak? When fitting line profiles of mean atmosphere models, a microturbulent ingredient 
was used in the past and sometimes interpreted as evidence for small scale turbulence. This 
interpretation has become obsolete, however, now that detailed line profile calculations from 
the numerical simulations (Lites et al. 1989, Dravins and Nordlund, 1990) reproduce the 
observed line profiles with high accuracy. This is achieved already with a grid that does not 
resolve scales much smaller than an intergranular lane. This puts strong limits on the amount 
of hidden small scale flows that is present on the Sun but not in the simulations. 

The astrophysical simulations do show some small scale turbulence\footnote
{for figures illustrating this see {\tt
http:$//$www.pa.msu.edu$/\sim$steinr$/$research.html}
\hfill\break and {\tt http:$//$www.astro.ku.dk$/\sim$aake$/$ }}, 
but most of the broadening of the line profiles is due to the velocity field on 
the granular scale. While this is a fortunate circumstance for the simulation efforts, one 
may wonder about the causes, especially since certain forms of turbulence are seen in the 
laboratory case. One reason for the low level of turbulence in the surface layers of stellar 
convective envelopes is the fact that the vacuum outside a star does not support any stress, 
so that little shear is produced in the surface boundary layer by the top boundary 
conditions. The Reynolds number of this shear is still very large, but its the amplitude is 
small compared with the mean flow in the cells. Turbulence  generated locally is weak since 
the turnover time in this weak shear is longer than the time during which the flow is exposed 
at the surface: local turbulence generating processes do not have enough time to grow. 
Another reason is the extreme density 
stratification of the envelope. Since the upflows derive from deep layers, any fluid that 
reaches the radiating surface has expanded by a large factor. This expansion also dilutes any 
vorticity that might have been present. Since the upflows contain little small scale 
vorticity to begin with, the level of turbulence will be small in the upflowing material 
exposed at the surface. This differs strongly from the laboratory case, where turbulence near 
the surface is created locally due to the no-slip boundary condition as well as being 
advected from the interior of the cell.

\subsection{Radiation}
The second important difference with laboratory convection is that the surface boundary layer 
is determined by radiation rather than thermal conduction. To get its structure 
quantitatively right the physics of radiative transport has to be included at a fairly high 
level. The dependence of the opacity on the thermodynamic variables and on frequency have to 
be taken into account, and radiative transport has to be solved with sufficient accuracy in 
three dimensions. All these factor contribute to the `bottom line', the effective surface 
temperature (if the entropy at the base is given, or vice versa). Models of stellar 
convection which ignore any of these factors can get the surface 
temperature right only by accident or by calibration of a free parameter. This is in stark 
contrast to the laboratory case, where the effect of thermal diffusion can be measured by a 
single dimensionless quantity, the Rayleigh number, which is known in advance. The second 
role of the radiative processes is of course that they provide all our 
observations. In this role they are crucial by providing the interface at which essential 
aspects of the models can be tested against reality. 

Further important factors not present in the laboratory case include the physics of 
ionization, which influences the equation of state, and the viscosity. Whereas in the 
laboratory the viscosity as measured by the Prandtl number is of order unity, it is 
negligible (of the order $10^{-5}$ or less) in stellar envelopes. At $Pr\sim 1$ viscosity is 
an important factor in the boundary layers, but at low $Pr$ its influence may well be 
different.

\section{State of the art in numerical simulation}

\subsection{surface convection}

\begin{figure}
\mbox{}\hfill\epsfysize16cm\epsfbox{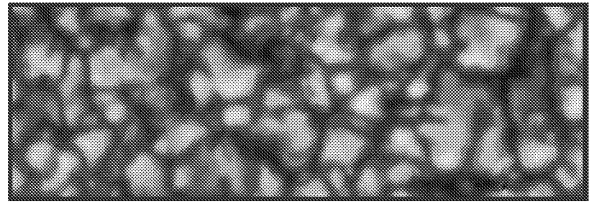}\hfill\mbox{}
\caption[h]{\label{comp}
Synthetic images from models of the solar surface, compared with high resolution observations 
from the Swedish Solar Observatory on La Palma. The upper two panels are as calculated from 
the simulations, the middle two after application of a point spread function (PSF) combining 
the influence of telescope and atmosphere. This PSF reduces the intensity and Doppler 
velocities to the same amplitude as the observations. Size of each panel about $18\times6$Mm. 
From Nordlund and Stein (1996).}
\end{figure}

Figure \ref{comp} shows the level of accuracy currently reached by the best three-dimensional 
simulations. At the highest numerical resolution, the patterns produced by the simulations
(properly degraded to match the observational conditions) match the best observations 
so well that the difference is hard to tell. This turns out to be one of the stiffest 
tests, as the human eye is quite sensitive to small qualitative differences in patterns. 
Quantitative agreement with the observations had already been achieved at much lower 
resolution. Values for the entropy $S_{\rm b}$ of the convective envelope sufficiently 
accurate for stellar evolution can already be obtained with $64^3$ simulations, or even with 
two-dimensional simulations (Ludwig et al. 1996). 

\begin{figure}
\mbox{}\hfill\epsfysize7cm\epsfbox{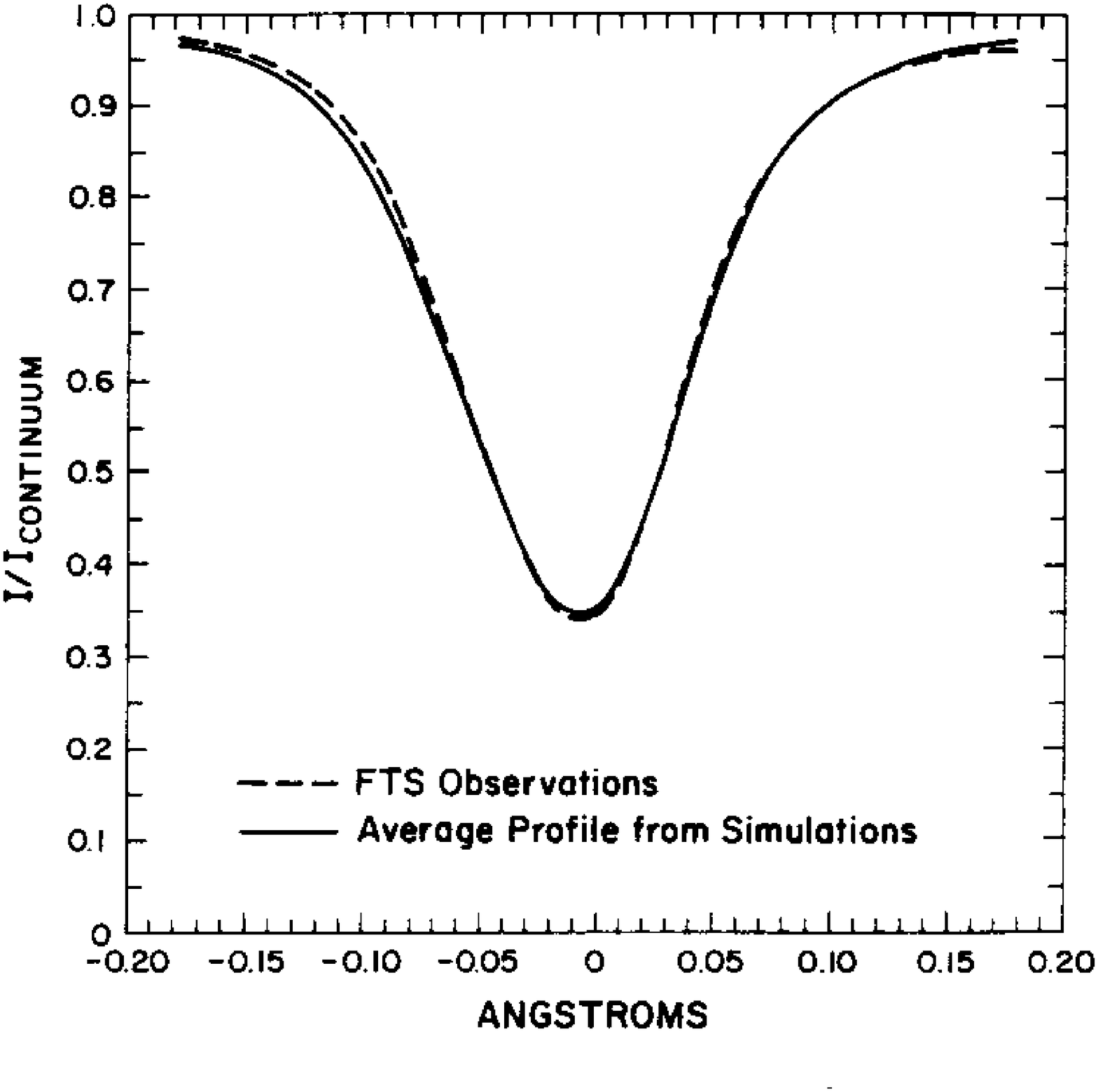}\hfill\mbox{}
\caption[h]{\label{line}
Comparison between synthetic and observed profiles of a popular iron line in the solar 
spectrum ($\lambda$6302.5). About half of the width of the line, and all of its asymmetry and 
wavelength shift result from the details of the velocity and temperature fields. The level of 
agreement is as good as can be expected given the spatial and temporal variability of the 
line. From Lites et al.\ (1989).}
\end{figure}

The average profiles of lines in the 
stellar spectrum are reproduced accurately with similar resolutions. About half of the width 
of the profiles is due to Doppler broadening and the details of the profiles are a sensitive 
measure of the velocity and temperature field at the surface. The lines have different shapes 
depending on their strength and temperature sensitivity, so that reproduction of a 
significant number of lines of different strength and sensitivity is a tight test of the 
correctness of the velocity and temperature fields. The agreement with observations (Nordlund 
1985a,b, Lites et al.\ 1989) is even close enough to detect minor misprints in tables of the 
laboratory wavelengths (Scharmer, 1987, private communication). An example is shown in fig. 
\ref{line}. About half of the width of this line is due to the convective velocity field. It 
is asymmetric and has a net blue shift due to the correlation between the velocity and 
temperature fields. These asymmetries depend on $T_{\rm eff}$ and  $g$ and are different for 
other stars. Dravins and Nordlund (1990) show that for the well observed nearby stars 
$\alpha$~Cen~A and Procyon the agreement between simulation and observation is also 
excellent.

This level of agreement is obtained only when the known input physics is implemented with 
sufficient care. This includes, in particular, the opacities and their frequency dependence, 
proper treatment of the radiative transport problem, the thermodynamic properties of the 
partially ionized gas, and appropriately formulated open boundary conditions at top and 
bottom of the simulation to minimize boundary artefacts from contaminating the solutions. 
Many simulations exist in the literature in which quantitative reproduction of stellar 
conditions was not the main aim (e.g. Cattaneo et al. 1991). Though these often reproduce 
important features qualitatively, they also 
contain artefacts that confuse the interpretation when application to stars is attempted. 
Several groups of workers have now been able to reproduce the basic results of Nordlund and 
co-workers, and substantial agreement exists on the accuracy with which the input physics 
must be treated (Steffen et al. 1989, Steffen and Freytag 1991, Solanki et al. 1995, 
Atroschenko and Gadun, 1994).

The solar p-mode oscillation frequencies show  small systematic differences with respect to 
standard envelope models. The dependence of these differences on wave number and frequency 
show that the origin of the discrepancy lies in the uppermost layers. From the numerical 
models Nordlund and Stein (1996) conclude that the difference is due mostly to `turbulent 
pressure': the effect of the kinetic energy of the flows on the mean stratification and on 
the mean compressibility of the fluid. The strong inhomogeneity of the flows also contributes 
by modifying the mean propagation characteristics of the fluid. Excitation of the p-modes in 
the right frequency range and with realistic amplitudes has also been demonstrated with the 
numerical simulations (Stein and Nordlund, 1991), though identification of the precise 
mechanism(s) of excitation still requires more study.

At this point, it is appropriate to stress the significance of what has been achieved.
Starting with nothing more than the fluid equations, the equation of radiative transfer, and 
(a large amount of) known atomic physics, {\em ab initio} simulations have reached agreement 
with observations to the level apparent in figures \ref{comp}, \ref{line}. This is a rare 
feat in astrophysics. In terms of programming and computation resources it has been a 
demanding effort, comparable in scale to that of the astrophysical opacity programs.

\subsection{deeper layers}
The situation is much less favorable for models of convection in the deeper layers. I 
stress again that this is largely irrelevant if the aim is to calibrate the mixing length 
parameter in stellar evolution, or to reproduce surface observations since these depend only 
on the upper few thousand km of the convection zone. But the flow in the deeper layers 
influences other observables such as the differential rotation. Convincing simulations of the 
deeper layers will be much harder to do. The main reason is the extreme stratification. A 
flow that descends from the surface is compressed by a linear factor 100 (if the compresssion 
is isotropic). If the mixing of downflows with their surroundings has to be followed, even 
smaller length scales must be resolved. For this reason, it is presently not possible to 
predict the width, velocity and entropy contrast of `typical' downflows as they reach the 
base of the convection zone, since these depend critically on the amount of mixing. A second 
problem is a disparity of time scales. The thermal time scale of the convection zone is 
$10^5$ yr, compared with typical flow timescales of the order 1 month in the lower convection  
zone and 10 min at the top. Covering these widely varying length scales and time scales is 
not possible 
numerically, some simplifications are necessary to treat the deeper layers. This has been 
done in the past by replacing the upper layers by a boundary at a greater depth (e.g. 
Glatzmaier, 1985a,b). The results showed only modest agreement with observations (of 
differential rotation and large scale flows for example). From what we know about the physics 
in the upper most layers, this is not surprising, since a fixed boundary at some depth is a 
very poor representation of the forest of narrow cool threads produced at the solar surface. 
This is probably the reason why such simulations produce large scale flows with amplitudes 
that are about two orders of magnitude larger than observed. It is not clear at present how 
these limitations can be overcome.

\section{Rethinking}
\label{reth}
Though the new picture of stellar convection that has emerged is physically simple and, after 
the fact, intuitively rather obvious, it conflicts somewhat with several currently standard 
ideas and physical explanations of processes in a convective envelope. In some cases these 
old views can be translated in a simple way in terms of the new physical picture, but in  
other cases it appears that more serious thinking is required.  
\subsection{The spectrum of length scales at the surface}
\label{lsc}
The velocity and temperature fields at the solar surface have been measured in detail in many 
studies. In the published interpretations a strong desire may be noted to make these 
measurements agree with turbulence models, in particular, of course, with a Kolmogorov 
spectrum. Usually it was realized in these studies that assumptions of isotropy and 
homogeneity are very strongly violated by the fact that the scale height of the solar 
atmosphere is {\em smaller than any of the length scales that can be directly measured}, and 
that the flow patterns observed do not naturally suggest interpretation in terms of 
Kolmogorov turbulence. But it was felt that while this `additional physics' would complicate 
the picture, things would somehow sort themselves out so that the end result would be  
similar to a Kolmogorov spectrum and that any observed deviations from this spectrum could be 
regarded as second order effects. 

The most important implication of this view was that it assumed as {\em given} that the 
observed spectrum is the result of a cascade process in which kinetic energy 
is put in at a scale much larger than that of the granulation, and that the smaller scales, 
including granulation, would result from this in the form of an intertial range. From what we 
know now, this is not a meaningful interpretation. The cooling at the surface that drives the 
flows acts primarily on the granulation scale. The power seen on smaller scales is due, in 
part, to the fact that granules change size, some growing while others are squeezed away by 
growing neighbors. In part, the small scales in a power spectrum come about just because the 
boundaries between cells and lanes in the granular pattern are very sharp. These statements 
can be made with confidence because the velocity and intensity patterns and their evolution, 
of which the spectrum of length scales is only a mathematical derivative, can be observed in 
detail directly. These effects have nothing to do with a turbulent cascade. 

On scales larger than granules, the simulations show that the dominant effect is the merging 
of intergranular downflows with depth (Stein and Nordlund, 1989), causing the upflows to form 
larger horizontal scales as well. (The effect is also seen in numerical simulations of the 
laboratory experiments by Kerr, 1996). The upflows can not be the cause for the larger scales 
themselves, since they are neutrally buoyant. They do not drive any flows since they can not 
convert a buoyancy force into kinetic energy. The formation of the larger scales can be 
understood by looking at the mean density stratification. At the surface, the dense 
downdrafts contribute more to the horizontal mean than they do further down: because of the 
density stratification, they occupy a decreasing fraction of the volume. Comparing this mean 
stratification with the neutrally buoyant upflows, one then finds it to be unstable to 
overturning motions. The resulting larger scale flows 
cause the small scale downdrafts to merge with increasing depth. The larger scales can thus 
be regarded as a collective instability of the dense downdrafts. Like the small scales, they 
are driven by cooling at the surface, rather than by buoyant upwelling from below, since 
their cause lies in the dynamics of fluid that has just cooled at the surface. In a 
traditional mean-gradient model, one would have argued that the large scales are formed 
because the mean entropy gradient is also unstable on large length scales. On the level of 
mean models these descriptions are of course equivalent. The description in terms of the 
downdrafts, however, makes clear why the large scales do not carry an additional heat flux, 
just as observed on the Sun. This is hard to understand in terms of large scales created by 
thermal buoyancy in a mean gradient.
That the granules provide the driving force for the larger scale flows is also illustrated 
nicely by a numerical experiment (Nordlund, private communication). Starting with a 
simulation which had reached its statistically steady state, the cooling at the surface was 
switched off, and the decay of the flows observed. Since no new downdrafts were generated, 
the flows at the top decayed on the local turnover time. The flows at each depth stopped as 
soon as the pre-existing downdrafts cleared the region above. Decay of large scale flows at 
the surface on a longer time scale, as would be expected if they were driven at a deeper 
level, was not observed.

Recapitulating, a turbulent cascade may produce approximate power law spectra of length 
scales, but other things can also produce such spectra, and this appears to be the situation 
at the solar surface. 

\subsection{Turbulent diffusion} 
In the absence of detailed knowledge of the convective flows in a stellar envelope, a simple 
model for the transport properties of the flows was to assume that they could be modeled by 
turbulent diffusivities, a turbulent viscosity for momentum and a similar quantity for the 
transport of heat. The mixing length model for the structure of stellar envelopes can be 
formulated in such terms. Justification for 
the model came from theories of homogeneous turbulence. Now that we know how extremely 
nonlocal convection is, it is appropriate to ask how relevant a picture based on 
a local turbulent diffusion is. Can we still use turbulent diffusion models for, say, 
explanations of the solar differential rotation? It may turn out that in a sufficiently 
coarse sense the transport of momentum in the convection zone can still be described in this 
way, but there is no particular reason to assume this any more, knowing, as we do now, 
that at high $Ra$ upflows and downflows can cross a convecting layer with little mixing 
between them.

\subsection{Sunspot heat flux blocking}
A second example of a problem in which the turbulent diffusion model has been used is the 
question what happens to the heat flux that is `blocked' at the surface by a sunspot. Where 
does this blocked heat flux reappear? A succesful answer (Spruit 1982a,b, Foukal et al. 1983) 
was found using 
a diffusion model for the flow of heat in a convective stellar envelope. Without going into 
the details here, these models predicted that practically all the `missing' heat flux is 
stored in the convection zone (which has a very long thermal time scale). These 
calculations successfully reproduced the absence of brightenings surrounding spots. How is 
the blocking problem explained without appealing to a turbulent diffusion? 

Below the spot 
(modeled as a region of reduced heat 
transport efficiency extending to some depth below the surface) the upflows have exactly the 
same temperature as upflows in the unspotted surroundings at the same level, namely that 
given by the entropy at the base of the upflows. In this sense, there is no `pile up of heat 
below the spot'. Because of the reduced heat loss at the surface, however, the downflows 
below the spot will be less vigorous. The unspotted surface notices nothing of the spot's 
presence (except for an extremely narrow ring where lateral radiative exchange takes place, 
and except for the presence of a moat flow, see below). It continues to cool upwellings into 
downdrafts as before, since the conditions in the upflows have not changed. Thus, the absence 
of bright rings around spots has an even simpler explanation than in the diffusion model. 

The spot is just a region at the surface where less heat is radiated away, and this is 
independent of the depth of the spot below the surface (in contrast to the diffusion model, 
where the spot has to extend to a minimum depth of 1000 km for the explanation to work). 
Still, one may wonder what happens to the amount of heat generated in the solar interior that 
now fails to be emitted at the surface. This part of the problem is the same as in the 
diffusion model: the imbalance causes a secular increase of the entropy in the entire 
convection zone until a new thermal equilibrium is reached (in detail: reheating of the cool 
downflowing fluid at the base of the convection zone to the entropy of the upflowing fluid is 
now quicker because the downflows below the spot are less vigorous. The imbalance causes the 
entropy of the upflows to increase). Because of the very long thermal time scale of the 
convection zone ($10^5$ yr), the effect is negligible on observable time scales. In a steady 
state, when the average number of spots does not change, the convection zone does not heat 
up, because its mean temperature is higher than it would be without spots (Spruit and Weiss, 
1986). Episodes of larger than average spot coverage cause heating, those of less than 
average spot coverage cause cooling on this time scale. 

\subsection{Moat circulation}
In the diffusion model, the horizontal flow away from a spot observed as the `moat' has a 
simple explanation, since the convection zone is hotter than average below the spot in this 
model, causing an upflow by thermal buoyancy. In our new view of the convection zone, the 
explanation does not work any more in this simple form, since there is no upward buoyancy 
anywhere. Still, the explanation is almost the same as before, namely buoyancy due to a 
reduced density below the spot. Downflows are cooler and denser than the surrounding 
upflows, so that the {\em mean density} on a horizontal surface is greater than the density 
in the upflowing fluid. Below the spot this effect is less strong than outside the spot, 
since less cooling has taken place at the surface. This difference in mean buoyancy of the 
fluid below the spotted and unspotted areas drives a circulation as before. The difference in 
explanation may sound a bit pedantic, since one could also have said that the mean 
temperature below the spot is higher than outside because of the weaker downflows (cf 
dicussion in section \ref{lsc}). I find the new explanation more satisfactory since it 
appeals more directly to the physical cause of the circulation, namely an imbalance in the 
fluid density. Note that in this explanation, the moat circulation does not carry extra heat 
to the surface: the mean temperature on a given horizontal level at some depth is higher in 
the circulation, but this increase is entirely caused by fluid that is moving downward!

\subsection{Erupting magnetic flux tubes}
The major advance in the past decade in our understanding of the solar cycle has been the 
realization that the magnetic fields we see at the surface started their lives at the base of 
the convection zone with strengths much larger than expected on the basis of equipartition 
with the convective flows. Numerical simulations of the eruption process (D'Silva and 
Choudhuri 1993, D'Silva and Howard 1993, Caligari et al. 1995) have demonstrated convincingly 
that the observed heliographic latitude of eruption and tilt of active region axes require 
strongly buoyant fields, with strengths of the order $60\,000$G. In these calculations, the 
convection zone has been treated with a mixing 
length model. The buoyancy of the fields is due both to the magnetic pressure and the thermal 
buoyancy acquired by traveling upward in the mean superadiabatic gradient. In the new view, 
however, there is no thermal buoyancy in anything moving upward. To replace the mixing length 
model, one would have to compute the eruption process in a constant entropy stratification. 
The mean upward velocity of the fluid surrounding the magnetic field should be taken into 
account in such a model; it replaces the effect of the thermal buoyancy in the superadiabatic 
mixing length model. It is not clear at present what, if any, the difference would be for the 
flux eruption calculations, but this could be checked easily.

\subsection{Lithium depletion}
\label{li}
The abundance of Lithium at the solar surface shows that the material of the convection zone 
has been exposed to temperatures at which Li burns. These temperatures are reached only at a 
depth of about half a scale height below the base of the convection zone. This shows that 
some form of weak mixing exists below the convection zone, such that the depletion time scale 
for Li is about $10^9$yr, shorter than the age of the Sun, but very much longer than the 
mixing time scale in the convection zone itself. Among the possibilities are 
rotation-induced shear instabilities (Zahn 1994), mixing by internal gravity waves (Press 1981, Garc\'{\i}a L\'opez and Spruit 1991), and some form of convective 
`overshooting' (e.g. Freytag et al.\ 1996). The main difficulty with overshooting in its 
traditional form is that mixing 
into the stable layers is possible only by flows of sufficient speed to overcome the stable 
stratification below the convection zone. Flows of lower energy only induce internal gravity 
waves. Whatever mixing is caused must be very mild compared with mixing in the 
convection zone, since not all Li has burned. Mild flows on the other hand are not able to 
penetrate significantly into the stable layers, and are not able to change the stratification 
in these layers. This conflict between depth of penetration and its required mildness is 
difficult to achieve with known forms of convective overshooting. In the new picture of 
convection, new possiblities arise for explaining the Li depletion (and other abundance 
anomalies like the $^{13}$C anomaly in giants), of which the following is a speculative 
example.

The downflows are cooler and have much higher speeds than in mixing length models, leading to 
potentially much more `interesting' mixing at the base of the convection zone (Roxburgh, 
1984). In order to penetrate a fraction of a scale height, however, the flows must reach a 
reasonable fraction of the sound speed. This is unlikely because of the drag between the 
downdrafts and their surroundings. With a standard estimate of the aerodynamic drag the 
penetration depth is still negligible. One might speculate, however, that in a small number 
of the downdrafts the entropy acquired at the surface would be nearly conserved, in spite of 
heating by thermal diffusion and mixing with the surroundings. This might be the case if some 
downdrafts collect into sufficiently large clusters. Irrespective of the velocity with which 
they arrive, these would then embed themselves into the stable layers at the depth where they 
are neutrally buoyant, corresponding to their entropy. An entropy contrast corresponding to a 
density 
difference $\delta\rho/\rho \approx 0.1$ would be sufficient to reach the Li burning layer. 
Even if this mechanism works, the observed level of depletion implies that only a very small 
fraction of the downdrafts makes it to this depth. Making a model that predicts this small 
number convincingly may not be so easy.

\section{Conclusion}
The numerical simulations of stellar surface convection pioneered by Nordlund represent one 
of the most important advances in stellar physics of the past two decades. They eliminate 
major sources of uncertainty in stellar models, but are at least as important because of the 
new qualitative insights they provide. While the astrophysical situation is different from 
laboratory convection in several essential ways, there are siginificant similarities which 
can be understood in a common language. Both lead to a new way of looking at convection. 

Several aspects of this new picture are easy to visualize. The main difficulty in 
accepting them is probably that they require, at some level, a break with the paradigm `large 
length scales equals large Reynolds numbers equals cascades of (local, approximately 
homogeneous) turbulence'. Though some of the concepts in this paradigm can be stretched to 
cover new evidence (e.g. by terms like `hard turbulence'), this practice also makes it harder 
to accept the consequences of key findings like the extremely nonlocal nature of the flows 
even on small scales.

The examples described in section \ref{reth} give an idea which observations are affected by 
the current understanding of stellar convection, and how this might possibly lead to new ways 
of explaining old phenomena. It may take a while before such views will eventually displace 
currently familiar conceptual tools such as turbulent diffusivities and buoyant bubbles 
moving in a mean superadiabatic gradient. These concepts now may still look comfortable in 
their familiarity, but may be just as misleading as the celestial spheres revolving around 
the earth were before the heliocentric view became accepted. It is the physical rigor of the 
direct simulations, as much as the mass of observables to be explained, that will necessitate 
a change of the theoretical reference frame.

%

\acknowledgements
I thank Drs. {\AA}. Nordlund,  R. Kerr, R. Stein and H.-G. Ludwig for discussions and for 
comments on an earlier version of the text.


\end{document}